\documentclass[aps,pra,twocolumn,showpacs]{revtex4-1}
\usepackage{graphicx}
\usepackage{amssymb,amsfonts,amsmath}
\usepackage[colorlinks=true,citecolor=Cerulean,linkcolor=RubineRed,urlcolor=Cerulean]{hyperref}
\hypersetup{breaklinks=true}
\usepackage{graphicx}
\usepackage{color}
\usepackage[usenames,dvipsnames]{xcolor}
\usepackage{epstopdf}
\usepackage{units}

\newcommand{\Ai}{\mathrm{Ai}}

\newcommand{\Jeff}{J_\mathrm{eff}}
\newcommand{\Hlatt}{H_\mathrm{latt}}
\newcommand{\vmax}{v^\mathrm{max}}

\begin{document}

\title{Relativistic motion of an Airy wavepacket in a lattice potential}

\author{C.E.~Creffield}
\affiliation{Departamento de F\'isica de Materiales, Universidad
Complutense de Madrid, E-28040 Madrid, Spain}

\date{\today}

\begin{abstract}
We study the dynamics of an Airy wavepacket moving in
a one-dimensional lattice potential. In contrast to the usual case
of propagation in a continuum, for which such a
wavepacket experiences a uniform acceleration, 
the lattice bounds its velocity,
and so the acceleration cannot continue indefinitely. Instead, we
show that the wavepacket's motion is described 
by relativistic equations of motion, which
surprisingly, arise naturally from evolution under the standard non-relativistic
Schr\"odinger equation. The presence of the lattice potential allows
the wavepacket's motion to be controlled by means of Floquet engineering.
In particular, in the deep relativistic limit when the wavepacket's motion
is photon-like, this form of control allows it to
mimic both standard and negative refraction. Airy wavepackets held
in lattice potentials can thus be used as powerful and flexible
simulators of relativistic quantum systems.
\end{abstract}

\maketitle

\section{Introduction}
In 1979 Balasz and Berry \cite{berry} demonstrated that the Airy
function is a non-diffracting solution of the Schr\"odinger equation
of a free particle in one dimension. As well as
being non-spreading, this solution also has the unique property
of appearing to accelerate at a constant rate, even in the absence
of an external force. Following on from this result, a great deal of
work has been done both theoretically and experimentally to study
this type of behaviour, with a particular eye to harnessing its unusual
properties for applications such as particle manipulation 
\cite{ballistic,particle_clearing}, optical routing \cite{routing}, 
and microscopy \cite{microscopy}.
The first experimental observations of Airy wavepackets
were made in an optical system \cite{siviloglou}, making use of the
correspondence between the paraxial diffraction equation and
the Schr\"odinger equation. Later work showed how Airy wavepackets
of free electrons could be generated \cite{electron_airy}, and
more recently, it has been suggested to use systems of Bose-Einstein
condensates \cite{wolf,yuce_mod}, in order to generate self-accelerating
matter waves.
  
The majority of studies have concentrated on the continuum case.
Lattice systems, however, represent a fascinating arena to
investigate and make use of the dynamics of Airy wavepackets.
Such systems must clearly reproduce the continuum behaviour as 
the lattice spacing is reduced to zero, but can be expected to 
show novel features arising from the interplay between the spatial
discretization and the self-acceleration effect. 
In particular, systems of ultracold atoms held
in optical lattice potentials \cite{lewenstein}
are excellent candidates for studying these effects, 
due to their high degree of quantum coherence and their 
controllability \cite{jaksch}. Such systems have already been used
as idealized lattice simulators to emulate the quantum dynamics of
condensed matter systems, such as the Hofstadter butterfly 
\cite{aidelsburger,ketterle},
the direct observation of Bloch oscillations \cite{bloch_oscs},
and Veselago optics \cite{veselago_lens}.

In this work, we investigate the dynamics of an Airy wavepacket moving on
a tight-binding lattice. We will firstly see that the initial behaviour of
this system duplicates that of the continuum case, with the wavepacket
undergoing a constant acceleration. For longer times, however, this
acceleration reduces towards zero and the velocity of the wavepacket 
saturates to a maximum value. 
This is a direct effect of the spatial discretization,
since the speed of propagation in the lattice is bounded by the lattice
dispersion relation. As a consequence the kinematics of the system
is described very accurately by the formalism of special relativity,
with the lattice group velocity playing the role of the ``speed of light''.
Rather unexpectedly, this relativistic description emerges spontaneously
from the dynamics of the standard non-relativistic Schr\"odinger
equation. This contrasts, for example, with the behaviour of
Gaussian wavepackets in a tilted lattice; these are also subject to
a constant accelerating force, but instead undergo Bloch oscillation
\cite{zener,bloch}.
We then show how Floquet engineering \cite{floquet} allows us
to manipulate the system's dispersion relation, and thus control the
propagation of the wavepacket. This demonstrates
how using Airy wavepackets in lattice potentials is a
convenient and powerful method to simulate relativistic quantum systems.

\section{Airy wavepacket}

\subsection{Self-acceleration}
\begin{figure}
\begin{center}
\includegraphics[width=0.45\textwidth,clip=true]{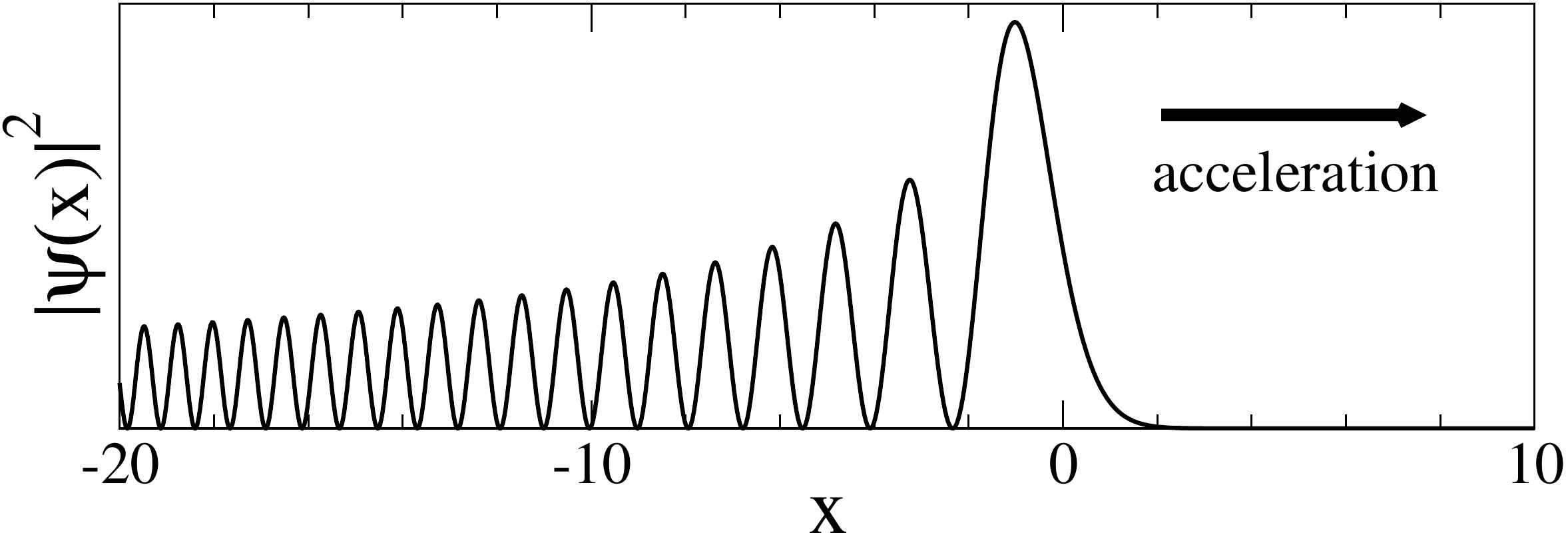}
\end{center}
\caption{Probability density of the Airy wavepacket given
in Eq. \ref{airy} at $t=0$.  For positive values of $x$ the probability 
density drops sharply with distance, while for negative $x$ the function
has a decaying oscillatory behaviour. The slowness of this decay 
(\ref{asymptotic}) means that the Airy function is not normalizable.
Under the action of the Hamiltonian in Eq. \ref{schrod}
this wavepacket will accelerate to the right at a constant rate.}
\label{initial_state}
\end{figure}

We begin by considering the Schr\"odinger equation for a particle
of mass $m$, moving in a one-dimensional system in the absence of
any external potentials
\begin{equation}
i \hbar \frac{\partial \psi}{\partial t} = - \frac{\hbar^2}{2 m} \ 
\frac{\partial^2 \psi}{\partial x^2} \ ,
\label{schrod}
\end{equation}
where $x$ represents a dimensionless spatial coordinate and $t$
is the corresponding time coordinate. For convenience we shall now
set $m$ and $\hbar$ equal to one.
As was shown in Ref.\cite{berry}, a solution of
Eq. \ref{schrod} is given by the Airy function
\footnote{This corresponds to the solution given in
Ref.\cite{berry} with the explicit parameter choice of $B / \hbar^{2/3} = 1$.}
\begin{equation}
\psi(x,t) =  \Ai \left( x - (t/2)^2 \right) \exp\left(
i \left( x t / 2 \right) - i t^3 / 12\right) \ ,
\label{airy}
\end{equation}
which can be readily verified by direct substitution.
We plot the corresponding probability density for $t=0$ in 
Fig. \ref{initial_state}. For $x > 0$, the Airy function
decays rapidly with the form
$\Ai(x) \sim \exp\left[ -2 x^{3/2} / 3 \right] / x^{1/4}$,
which produces the well-defined wavefront at the right of the wavepacket.
Conversely, for large negative values of $x$, the Airy function
has a decaying oscillatory form \cite{stegun}
\begin{equation}
\Ai\left( x \rightarrow - \infty \right) \sim
\frac{1}{ \sqrt{\pi} \left| x \right|^{1/4}}
\sin \left( 2 \left| x \right|^{3/2} / 3 + \pi / 4 \right) \ .
\label{asymptotic}
\end{equation}

From Eq. \ref{airy} it is clear that the probability density, 
$\left| \psi(x,t) \right|^2$ preserves its shape over time, and that 
it follows a parabolic trajectory $x(t) = t^2/4$, that is, it appears to
undergo a constant acceleration of $a = 1/2$ in the system of units we use. 
Although this non-intuitive result would appear
to violate the Ehrenfest theorem, this is not in fact the case. As can
be seen from Eq. \ref{asymptotic}, the Airy
wavefunction is not $L^2$ integrable, and so its centre of mass
is undefined. Consequently we cannot interpret the acceleration as the
response of the system's centre of mass to a force.
If instead, however, we focus our attention on some specific points of
the wavefunction, such as, for example, the maxima of the probability
distribution, we will indeed see the locations of these points moving 
along parabolic trajectories in the $x-t$ plane 
(see Fig. \ref{trajectory} for small values of $t$.)
For the remainder of the paper we will concentrate
on the motion of the first, and highest intensity, peak of the wavefunction,
which at $t=0$ is centered on $x \simeq -1.019$, and use its position to 
calculate the wavepacket's velocity and acceleration.
As this acceleration arises spontaneously, in the absence of any external
potentials, we shall refer to it as ``self-acceleration''.

\subsection{Normalization and diffraction \label{diffraction}}
As the probability density is not normalizable, a true Airy wavepacket
can clearly not be prepared in experiment. One option is to use an
aperture-limited version of the function, by simply truncating the spatial
coordinate to run over a large, but finite, range of values. 
Smoother aperture functions can also be used to 
render the wavefunction normalizable, for example,
an exponential function, $\psi(x, t=0) = \Ai(x) \exp \left( \gamma x \right)$,
where $\gamma$ is a small, positive constant.
The Fourier transform of this function is given by
\begin{equation}
\tilde{\psi} (k) \propto e^{- \gamma k^2} e^{i k^3/3} \ ,
\label{transform}
\end{equation}
where the cubic phase term arises from the Fourier transform of the
Airy function itself. This provides a particularly convenient way
to synthesize aperture-limited Airy functions
in optical systems \cite{siviloglou},
by simply imprinting a cubic phase on a Gaussian beam using a phase mask, and
then making an optical Fourier transform of the result. 
An analogous technique is
also possible for matter-wave optics \cite{electron_airy,wolf}  
to engineer the appropriate wavefunction in Fourier space, or alternatively
amplitude and phase masks can be used to generate the wavefunction
directly in real-space \cite{wolf}.

Truncating the Airy function has the consequence that the wavepacket
is no longer diffraction-free, and so it broadens with time. This distance, over
which the truncated Airy wavepacket approximately maintains its form, is 
known as the Airy zone \cite{airy_zone}, and its size reduces as the 
degree of truncation is increased. The majority of the results we report below 
were obtained by simply limiting the range of the spatial coordinate,
and in each case it was verified that the range of $x$ used was sufficiently 
large that the results were insensitive to it over the time-intervals
considered. The simulations were also repeated using an exponential
aperture function for various values of $\gamma$ as a further check of the
stability of the observed effects.

\subsection{Lattice Airy wavepacket}
Instead of allowing the particle to move in free space 
as in Eq. \ref{schrod}, we now impose
a lattice potential $V(x) = V_0 \cos^2 k x$. In ultracold
atom experiments, this can be conveniently done by superposing
two counter-propagating laser beams \cite{optical_lattice}
to create an optical lattice potential whose depth, $V_0$,
is proportional to the laser intensity.
For sufficiently deep optical lattices, the atoms will localize
in the potential minima, and can accordingly be described
in a basis of site-localized Wannier functions. 
In this case the dynamics of the atoms can be accounted for well
by retaining only the hopping matrix element $J$ that connects
a site to its nearest neighbours \cite{jaksch}, yielding
the lattice Hamiltonian
\begin{equation}
\Hlatt = -J \sum_{j = -M}^{N} \left( a^\dagger_{j+1} a_j + \mathrm{H.c.} 
\right) \ .
\label{lattice_ham}
\end{equation}
Here $a_j (a_j^\dagger)$ are the usual bosonic annihilation (creation)
operators acting on lattice site $j$. The hopping
amplitude $J$ is related to the parameters of the continuum
Hamiltonian as $J = \hbar^2 / \left( 2 m \Delta x^2 \right)$, where $\Delta x$
is the lattice spacing which relates the $x$-coordinate to the
lattice site, $x = j \Delta x$. Henceforth we will use $J$ as the
unit of energy and frequency, and 
measure time in units of $J^{-1}$.

\section{Results}

\subsection{Motion of the lattice Airy wavepacket}
\begin{figure}
\begin{center}
\includegraphics[width=0.40\textwidth,clip=true]{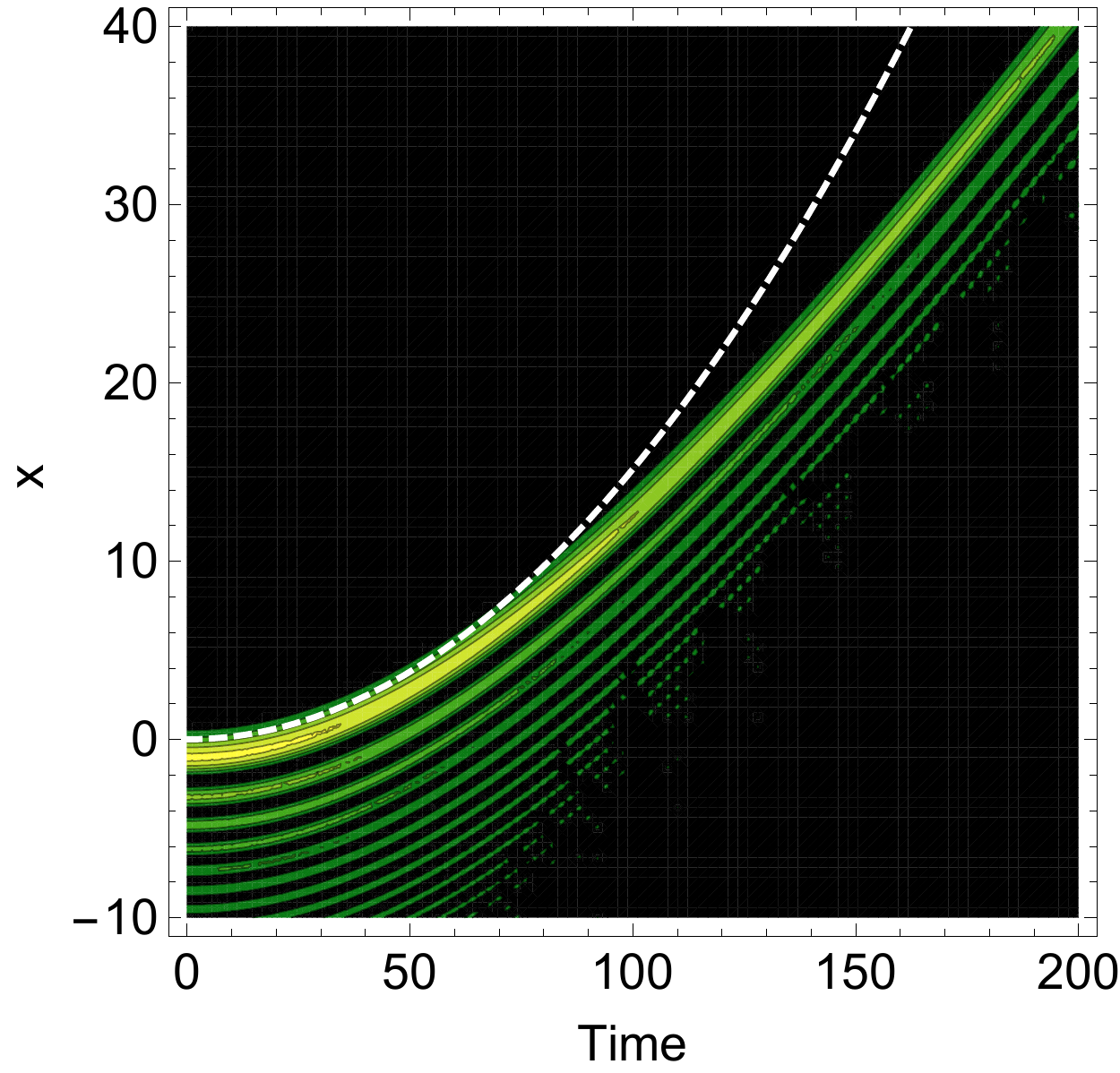}
\end{center}
\caption{Trajectory of an Airy wavepacket on a lattice. The system is
initially prepared in an Airy state, and evolves under the action 
of the lattice Hamiltonian (\ref{lattice_ham}). Initially the peaks in
the wavepacket move along parabolic trajectories corresponding to
a uniform positive acceleration, marked by the white dashed line.
At longer times, however, the acceleration of the wavepacket decreases
as it approaches relativistic speeds. Note that the amplitudes of the peaks
decrease with time due to wavepacket spreading; this occurs because
the Airy wavepacket is aperture limited (see Sec.\ref{diffraction}).
Lattice discretization, $\Delta x = 0.2$.}
\label{trajectory}
\end{figure}

In Fig. \ref{trajectory} we show the movement of a wavepacket initialized
in the state $\psi(x) = \Ai(x)$, where $x$ is a discretised
spatial coordinate with $\Delta x = 0.2$, under the time evolution
provided by the lattice Hamiltonian (\ref{lattice_ham}).
At $t=0$ we see the series of peaks in the probability density
produced by the $x$-dependence of the Airy function, with
the largest peak lying at $x \simeq -1$. For small values of $t$ ($t < 80$),
the peaks in the probability density move along trajectories which to
a good degree of accuracy are parabolic. We emphasise that this effect is
occurring in the absence of any accelerating potentials, and
so corresponds to the wavepacket undergoing a constant
self-acceleration.

For longer times, however, the movement of the peaks clearly begins to
deviate from the parabolic behaviour, with their location showing
a slower, linear dependence on time. This contrasts with the behaviour
of the continuum Airy wavepacket, which would continue to accelerate
indefinitely. To see the reason for this discrepancy
we can first note that the lattice Hamiltonian (\ref{lattice_ham}) can
be straightforwardly solved in momentum space
\begin{equation} 
\tilde{H}_\mathrm{latt} = -2 J \cos k \ \sum_k a_k^\dagger a_k \ ,
\label{kspace_ham}
\end{equation}
giving the dispersion relation $E(k) = -2  J \cos k$.
For small $k$ this reproduces the dispersion relation of a
free particle $E_\mathrm{free} = k^2 / 2$, but unlike
the free particle case, $E(k)$ is limited to a finite range. 
As a consequence, the group velocity, $v_g = 2 J \sin k$, 
has a maximum value of $2 J$, and so the system can 
only support excitations up to this maximum velocity $\vmax$. 
This is a specific instance of the Lieb-Robinson bound \cite{lieb}.
This motivates us to employ a relativistic description of the
system, with this maximum velocity playing the role of the speed
of light.

\subsection{Relativistic description}
There is a frequent misconception that the case of a body subject
to a constant acceleration cannot be treated within special
relativity, but instead necessitates the use of general relativity.
This, however, is not the case; special relativity is
completely capable of describing such motion \cite{desloge}.
We wish to relate the kinematic quantities measured by
an observer at rest with respect to the laboratory (or lattice), with
those measured in a uniformly accelerated frame (the instantaneous
rest-frame of the Airy wavepacket).
If we denote the (constant) proper acceleration by $\alpha$, then it is
a standard textbook exercise \cite{french} to show that the
quantities measured in the laboratory frame are given by
\begin{eqnarray}
a(t) &=& \alpha / \left( 1 + \left( \alpha t / c \right)^2 \right)^{3/2} 
\label{accel} \\
v(t) &=& \alpha t / \sqrt{  1 + \left( \alpha t / c \right)^2 } 
\label{velocity} \\
x(t) &=& \left( c^2 / \alpha \right)
\left(  \sqrt{  1 + \left( \alpha t / c \right)^2 } - 1 \right) 
\label{space} \ ,
\end{eqnarray}
where $c$ represents the speed of light, and we have imposed the initial
conditions $x(0) = v(0) = 0$.

In the non-relativistic limit $\alpha t \ll c$ it can readily be seen that
these expressions reduce to the familiar results of Newtonian mechanics,
and in particular that $x = \mbox{\nicefrac{1}{2}} \ \alpha t^2$.
In general, we can cast Eq. \ref{space} in the more revealing form
\begin{equation}
\left( \alpha x / c^2 + 1 \right)^2 -
\left(\alpha t / c \right)^2 = 1 \ ,
\label{hyperbola}
\end{equation}
to show that the particle follows a {\em hyperbolic} trajectory in space-time
\cite{rindler}. 
The asymptotes of this trajectory are the two straight lines
$x = \pm c t$, which form the light-cone for this object.
The slopes of these lines bound the velocity of the wavepacket velocity 
(\ref{velocity}) such that it never exceeds $c$, but approaches it 
asymptotically with time.

\begin{figure}
\begin{center}
\includegraphics[width=0.45\textwidth,clip=true]{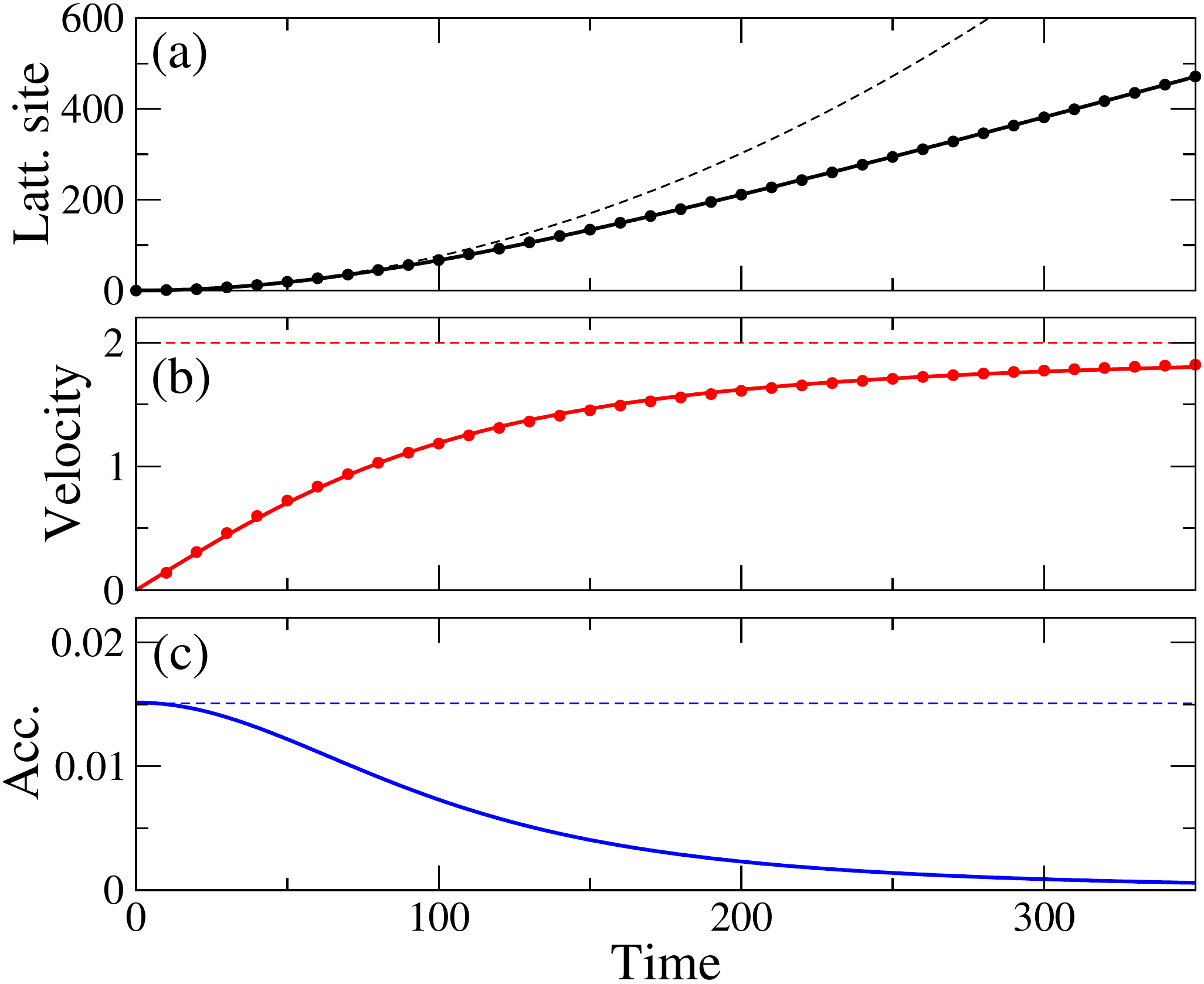}
\end{center}
\caption{Relativistic motion of the lattice Airy wavepacket,
obtained by a numerical simulation of the lattice system. \newline
(a) Symbols denote the lattice site occupied by the first peak of the 
wavepacket. Initially this position increases quadratically with time,
shown by the dashed line, but for $t > 100$ the increase becomes slower,
tending towards a linear rise. The solid line shows the relativistic
prediction (\ref{space}), where the values of $c$ and $\alpha$ are
obtained from a two-parameter fit to the data-points. The fit is seen
to be excellent.
(b) Velocity of the first peak of the lattice Airy function.
Symbols show that values obtained from the numerical time-derivative
of the data from the simulation, while the solid line shows
the theoretical result (\ref{velocity}). Initially the velocity
rises linearly, but flattens off as it begins to approach
the maximum lattice velocity $\vmax$, shown with the dashed line.
(c) Acceleration observed in the rest-frame of the lattice.
For small times the acceleration is constant, but drops as
the wavepacket enters the relativistic regime. The dashed line
indicates the proper acceleration, $\alpha$, which remains constant.}
\label{rel}
\end{figure}

In Fig. \ref{rel} we show the motion of
the first peak of the Airy wavepacket, for the same parameters as
in Fig. \ref{trajectory}. The data-points in Fig. \ref{rel}a are the 
motion of the peak obtained from the numerical time-integration of 
the system, while the solid line is the relativistic result for $x(t)$, 
where $\alpha$ and $c$ were obtained from a two-parameter fit of the
numerical data to Eq. \ref{space}. 
The agreement between the data and the fit is excellent.
The fit parameters are given in the first line of Table \ref{values},
and we can see that the obtained value of $c$ is indeed in good 
agreement with the theoretical value of $\vmax = 2 J$.
We also show in this figure the parabolic behaviour that would be
obtained in the non-relativistic case, which clearly emphasizes
the difference between the relativistic discrete case, and the
non-relativistic continuum behaviour.

Below in Fig. \ref{rel}b we compare the velocity of the main peak,
as calculated by the time-derivative of its location, with the
relativistic prediction of Eq. \ref{velocity}. Again the agreement
is seen to be excellent. The maximum lattice group velocity of
$2 J$ is also plotted, and the asymptotic saturation of the peak's
velocity to this value is clearly evident. Finally, in Fig. \ref{rel}c
we show the acceleration of the wavepacket, as measured in the
lattice rest frame. As the velocity of the wavepacket approaches
a significant fraction of $c$, this acceleration reduces smoothly
to zero.

\begin{table}
\begin{center}
\begin{tabular}{ |c|c|c| }
 \hline
$\Delta x$ & $c$ & $\alpha$ \\
\hline \hline
 0.20 & 1.90 & 0.0153 \  \\          
 0.15 & 1.85 & 0.0065 \ \\
 0.10 & 1.81 & 0.0020 \ \\
 0.05 & -- & 0.00025 \ \\
 \hline
\end{tabular}
\caption{Values of $\alpha$ and $c$ obtained by a two-parameter fit
to Eq. \ref{space} for various values of the lattice
spacing $\Delta x$. The acceleration $\alpha$ rapidly reduces as the lattice
becomes finer, and for the smallest value of $\Delta x$ the wavepacket
barely entered the relativistic regime in the time-span considered.
As a result the two-parameter fit was unstable, and $\alpha$ was simply
evaluated from a fit to a parabola.  In all other cases the value
obtained for $c$ is in good agreement with the theoretical
value of $\vmax = 2 J$.}
\label{values}
\end{center}
\end{table}

\subsection{Fitting and scaling}
In Table \ref{values} we show values obtained for $\alpha$ and
$c$ by fitting the numerical results for different
lattice spacings to Eq. \ref{space}.
Clearly, as $\Delta x$ is reduced, the value of the proper
acceleration decreases as well. For $\Delta x = 0.05$, the
smallest lattice spacing considered, the acceleration was
so small that the wavepacket did not enter the relativistic
regime over the time-interval considered. The length of
this time-interval was limited by the spreading of the wavepacket, 
produced by the aperture restriction of the wavepacket.
During this simulation time, the wavepacket
thus appeared to accelerate uniformly, which corroborates our
intuition that the behaviour of the discrete wavepacket should
approach that of the continuum case as $\Delta x \rightarrow 0$. 

\begin{figure}
\begin{center}
\includegraphics[width=0.45\textwidth,clip=true]{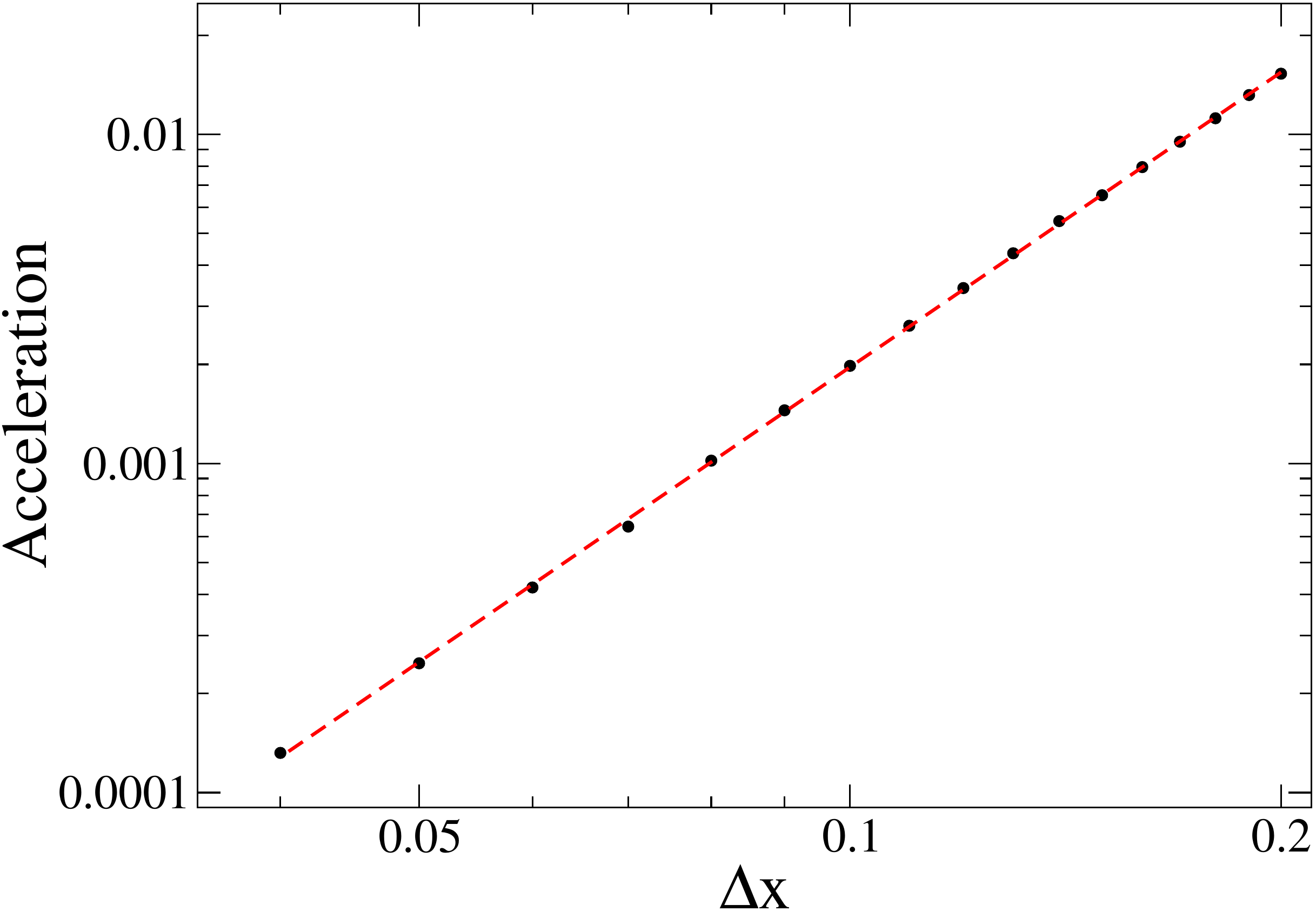}
\end{center}
\caption{Values of the proper self-acceleration $\alpha$, obtained by
curve-fitting the numerical results to Eq. \ref{space}. The scaling
of this quantity is described well by the power-law
$\alpha = 2 \Delta x^3$, shown by the dashed red line.}
\label{scaling}
\end{figure}

To put this observation on a quantitative basis, we show in
Fig. \ref{scaling} a logarithmic plot of the measured values of the
self-acceleration as a function of the lattice discretization.
The linear behaviour visible clearly implies a power-law 
dependence of $\alpha$ on $\Delta x$. To analyze this further,
and to investigate how the continuum result emerges as
the lattice spacing reduces to zero, we first recall that
in the continuum case
$x(t) = \mbox{\nicefrac{1}{2}} \ a t^2$, where for the Airy solution
we consider, $a = 1/2$. Writing the
time coordinate in the lattice units of $J^{-1}$, we find that
\begin{equation}
x(t) = \Delta x \ n(t) = \frac{1}{2} a \
\left( 2 \Delta x^2  t \right)^2 \ ,
\end{equation}
where $n(t)$ is the lattice site occupied by the peak
of the wavepacket. The self-acceleration of the wavepacket measured
in lattice-units is thus $\alpha = 4 a \Delta x^3$, and so we
should expect to see the power-law dependence 
$\alpha(\Delta x) = 2 \Delta x^3$. We plot this curve in Fig. \ref{scaling},
and indeed find that it describes the scaling of $\alpha$
extremely well.

\subsection{Bloch oscillation}
It is interesting to compare the case of the Airy wavepacket with
that of a particle in a tilted lattice potential, that is, a potential
which rises linearly in space, $V(x) \propto x$,
\begin{equation}
H_{\mathrm{BO}} = \Hlatt + V_0 \sum_j j n_j \ ,
\label{hamiltonian_BO}
\end{equation}
where $V_0$ is the difference in potential between neighboring
sites, and $n_j$ is the standard number operator.
Naively, one could interpret the lattice tilt as resulting
from the application of a constant force (since
$F = - \partial V / \partial x$), and so one would expect
the wavepacket to uniformly accelerate in the
direction of the tilt.
The presence of the lattice, however, complicates
this simple picture; although the wavepacket will initially
accelerate, it will also experience Bragg diffraction from the 
lattice potential. The result is that it will undergo
an oscillatory motion termed Bloch oscillation \cite{bloch,zener}. 

\begin{figure}
\begin{center}
\includegraphics[width=0.45\textwidth,clip=true]{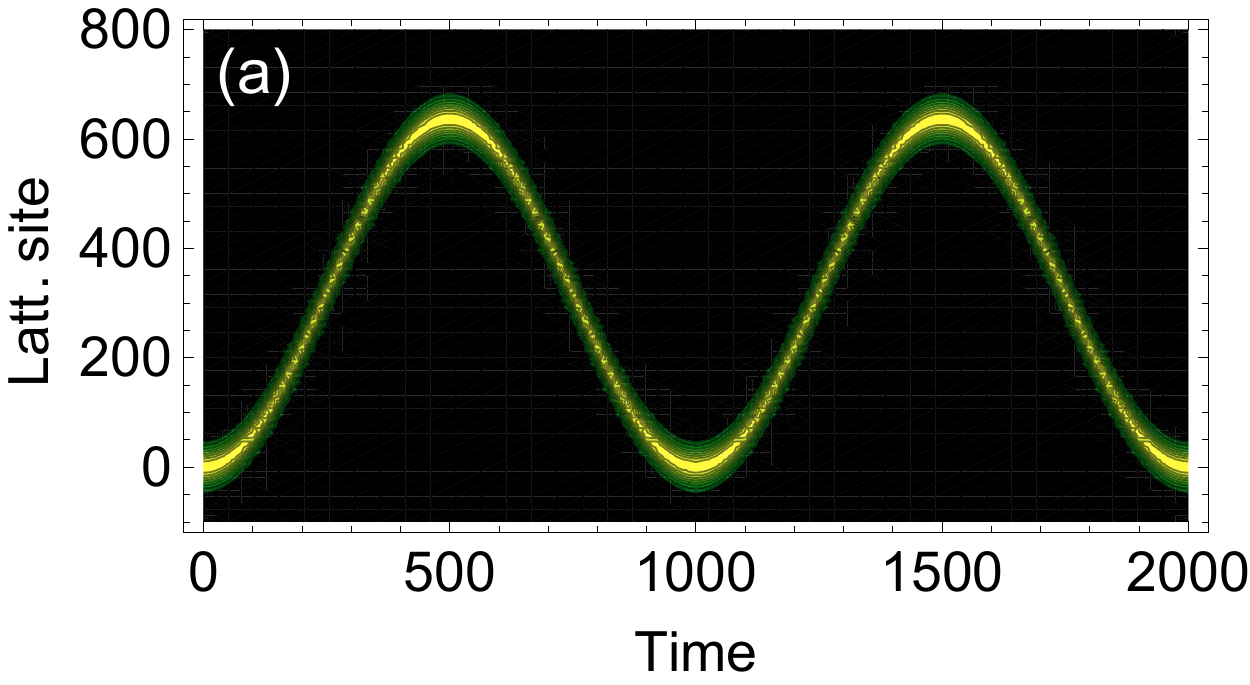}
\includegraphics[width=0.45\textwidth,clip=true]{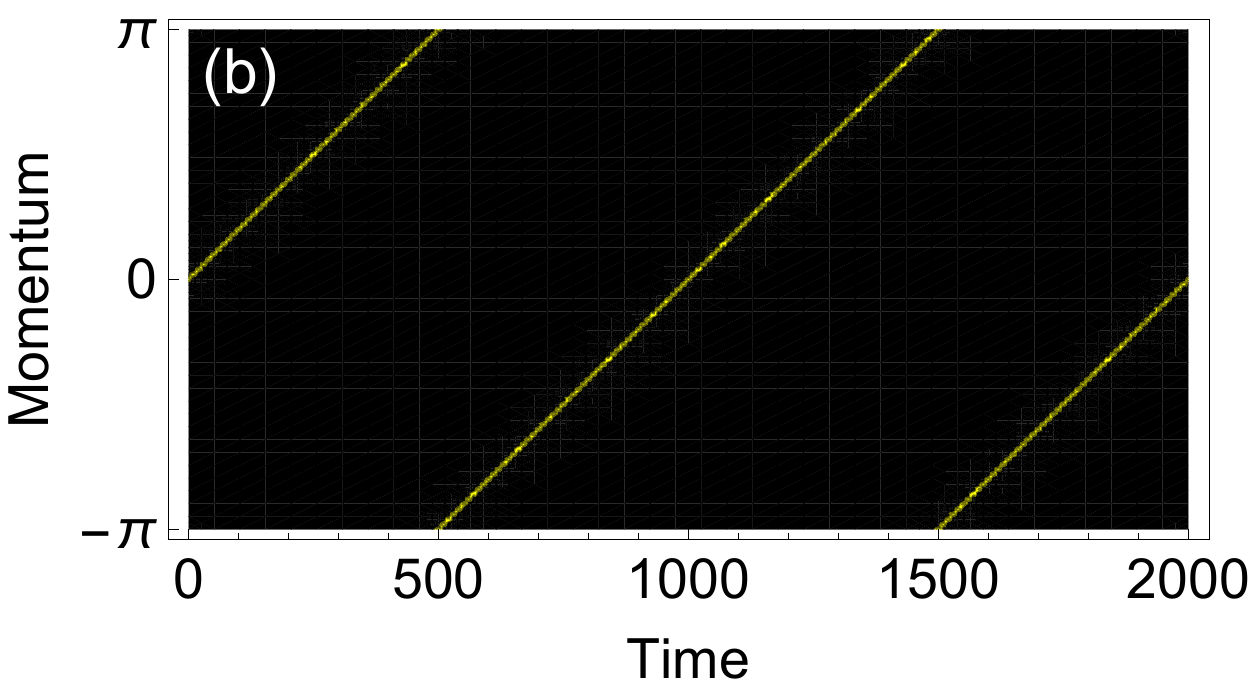}
\end{center}
\caption{Bloch oscillation of a Gaussian wavepacket in a tilted lattice.
(a) Probability density of the wavepacket as a function of time.
The wavepacket makes a slow oscillatory motion (Bloch oscillation)
described by Eq. \ref{com}.
(b) Probability density in momentum space, $ \left| \tilde{\psi}(k,t)|^2 \right|$
for the same system. The sharply-peaked distribution moves linearly
with time across the first Brillouin zone, crossing the boundary at
$t= 500$, corresponding to the reversal of motion of the
wavepacket in space.
Parameters of the system: $\Delta x = 0.2$, $V_0 = 2 J \pi / 1000$.}
\label{bloch}
\end{figure}

If the initial state of the particle is a well-localized wavepacket,
it is straightforward to show that the position of its centre of
mass is given by \cite{holthaus_bloch,dynamics_bloch}
\begin{equation}
x(t) = 2 \left( J / V_0 \right) \ \left( 1 - \cos V_0 t  \right) \ ,
\label{com}
\end{equation}
where for convenience we have set the initial condition $x(t) = 0$.
In Fig. \ref{bloch}a we show the numerical simulation
of a broad Gaussian wavepacket under the action of the
tilted lattice Hamiltonian (\ref{hamiltonian_BO}). The oscillatory
motion of the wavepacket is clear, the amplitude and frequency
of the oscillation being related to the size of the tilt. For
small values of $t$, we can make a Taylor expansion of Eq. \ref{com},
to reveal that it indeed begins to accelerate uniformly,
with $a = 2 J / V_0$. As with the discretised Airy wavepacket, this
acceleration reduces with time, but in contrast to the Airy case
it does not follow a relativistic form. Rather than 
asymptotically approaching the maximum lattice velocity, 
the wavepacket instead slows and eventually turns around, 
and begins propagating in the opposite direction.

The difference between this behaviour and that of the discretised Airy 
wavepacket is even clearer in momentum space. In Fig. \ref{bloch}b we show
the Fourier transform of the Bloch oscillation; the broad
Gaussian wavepacket in real space becomes a narrow Gaussian in momentum space. 
The effect of the lattice tilt is that the momentum distribution
shifts linearly with time, $k(t) = V_0 t$,
obeying the classical equation of motion $F = \partial k / \partial t$.
When the momentum reaches
the edge of the first Brillouin zone at $k=\pi$, it wraps around it
and reenters at $k = -\pi$, corresponding to the wavepacket's motion 
reversing. In contrast, the momentum distribution
of the Airy wavepacket does not alter with time (within the
Airy zone), even though we see the peaks of the wavepacket 
appearing to accelerate along hyperbolic trajectories.
\footnote{An alternative way of visualizing this is to see that
$\Hlatt$ commutes with $k$, and so the momentum of the Airy
wavepacket cannot change. Introducing the tilt breaks
this conservation law, and so the mean momentum is time-dependent 
in the case of Bloch oscillation.}.
This underlines the importance of taking care when discussing
the Airy dynamics. The Gaussian wavepacket obeys Ehrenfest's theorem, and
so we can consider its centre of mass to be accelerated by the
applied force. This is not the case for the Airy wavepacket.
Here the self-acceleration arises from quantum
interference effects, which depend on the entire form of the wavefunction.

\subsection{Wavepacket manipulation}
We have seen that the critical factor determining the propagation
of the lattice Airy wavepacket is the maximum velocity
of excitations in the lattice, $\vmax$. In turn this depends on the
hopping parameter $J$, indicating that if we can alter $J$ coherently
we will be able to finely control the trajectory of the wavepacket.
A powerful method to achieve this is provided by Floquet
engineering \cite{floquet}. In this approach, the lattice potential
is periodically driven in time, or ``shaken'', at a frequency much
higher than the other time-scales of the problem. In this high-frequency
limit, the full time-dependent Hamiltonian can be described by a
static effective Hamiltonian with renormalized parameters. In particular,
for the hopping Hamiltonian (\ref{lattice_ham}),
the tunneling is renormalized to an effective value $\Jeff$, and 
manipulating the parameters of the shaking permits the value
of $\Jeff$ to be  adjusted.

\begin{figure}
\begin{center}
\includegraphics[width=0.45\textwidth,clip=true]{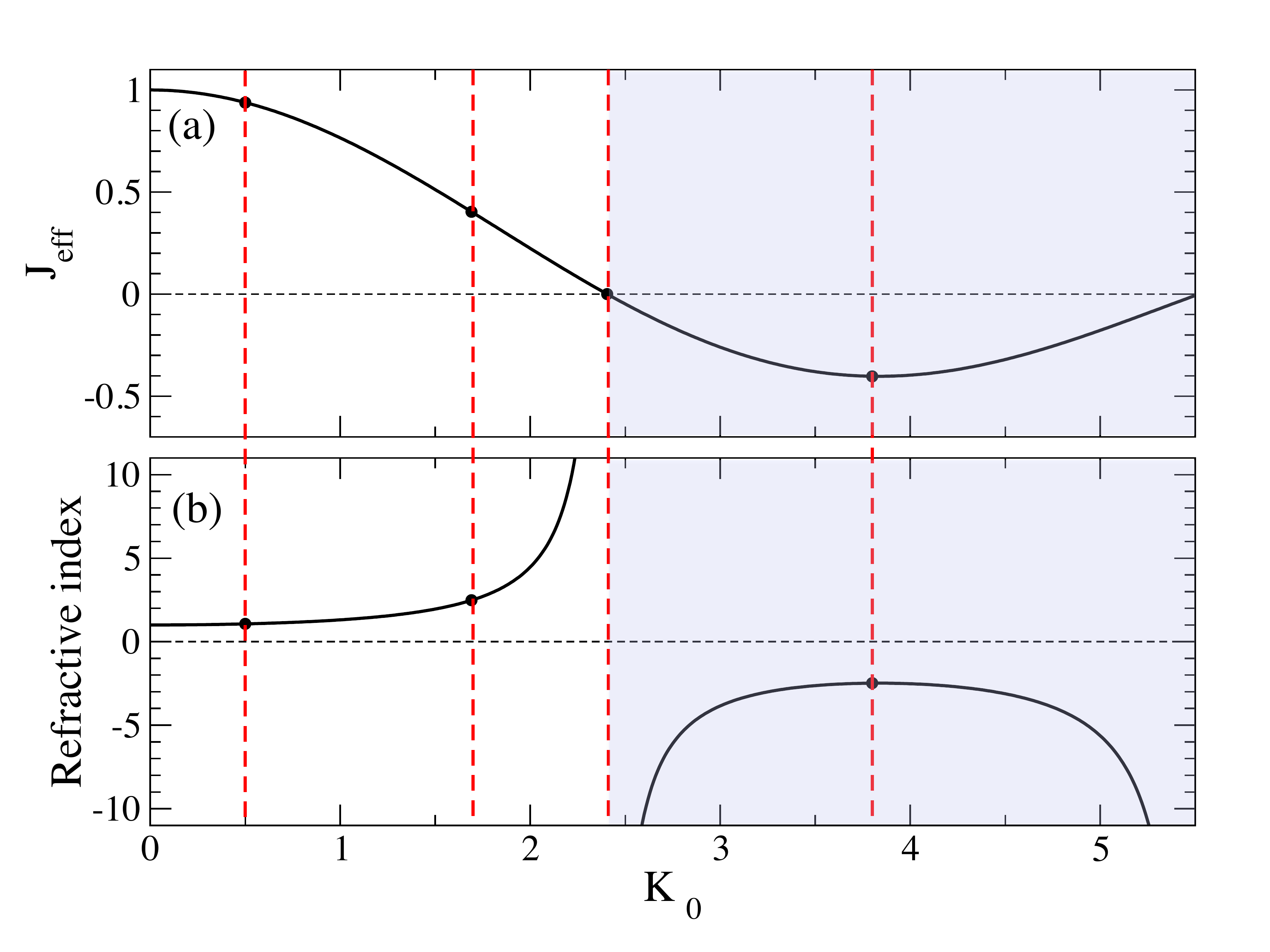}
\end{center}
\caption{(a) The effective tunneling, $\Jeff$, for a sinusoidally
driven lattice has a Bessel function dependence on the driving
parameter, given by Eq. \ref{bessel}.
The symbols mark the values of $K_0$ used to obtain the
results shown in Fig. \ref{driven}. For $K_0 = 0.5$, the
amplitude of the tunneling is slightly reduced from its
undriven value. At $K_0 = 1.691$,
$\Jeff = 0.403$, while at $K_0 = 2.4048$ the effective
tunneling vanishes. For $K_0 > 2.4048$, the shaded area, the effective
tunneling is negative; at $K_0 = 3.80$
the negative effective tunneling takes its maximum value.
(b) We denote the ratio of $\vmax$ in the undriven
system to $\vmax$ in the driven system as the refractive index.
As $K_0$ approaches the first zero of the Bessel function, the
refractive index increases, and diverges at $K_0 = 2.4048$.
In the shaded region, the refractive index is negative.}
\label{ref_index}
\end{figure}

We will consider the standard form of driving
\begin{equation}
H(t) = \Hlatt +  K \cos \omega t \  \sum_j j n_j \ ,
\label{driving}
\end{equation}
in which the potential has a sinusoidal dependence on time.
The behaviour of this Hamiltonian was studied in Ref.\cite{holthaus}
in the context of periodically-driven semiconductor superlattices,
and for the specific case of sinusoidal driving \cite{crossings}, 
the effective tunneling has the dependence
\begin{equation}
\Jeff = J \ {\cal J}_0 \left( K / \omega \right) \ ,
\label{bessel}
\end{equation}
where ${\cal J}_0$ is the zeroth order Bessel function
of the first kind.
For convenience we will use the notation $K_0 = K / \omega$
to denote the dimensionless argument of the Bessel function.
This form of driving has been used in cold atom experiments
\cite{lignier}, and the Bessel function dependence of the tunneling
has been directly observed \cite{creffield_pisa,guery}. The form of
$\Jeff$ is shown in Fig. \ref{ref_index}a. We can note that
at $K_0 \simeq 2.4048$, the first root of the Bessel function,
the effective tunneling vanishes. This produces the effect
known as CDT (``coherent destruction of tunneling'') \cite{cdt},
in which the tunneling dynamics of the system is completely
quenched. This effect has been used to induce the Mott transition
\cite{mott_theory,mott_expt}, and to control the motion of
quantum particles on lattices \cite{creffield_control}.

If we prepare the system as a lattice Airy wavepacket and allow it
to evolve freely, we have seen that it will undergo a relativistic
acceleration with its velocity approaching arbitrarily close
to the maximum lattice velocity $\vmax = 2 J$,
giving it a photon-like behaviour.
This can also be achieved by giving the wavepacket a kick,
by imprinting a phase on it of the form $\exp \left[ i \phi j \right]$,
where $j$ labels the lattice site, which imposes an initial
velocity of $v = 2 J \sin \phi$ on the wavepacket. By using 
a value of $\phi$ close to $\pi/2$ we can thus place the wavepacket
deep in the relativistic regime without waiting for it to accelerate
to this state from rest. Although the results reported in this section
were obtained by means of this phase-imprinting technique, we have verified
that the same results are obtained by allowing the wavepacket
to self-accelerate to this regime.

If we consider the propagation to be photon-like, with
$\vmax$ playing the role of the speed of light,
then controlling the magnitude of $J$ gives us control of a
quantity analogous to the refractive index. This will be the
ratio of $\vmax$ in the undriven system to the maximum
velocity when the lattice is shaken, which is simply given by
$J / \Jeff$. We plot this quantity in Fig. \ref{ref_index}b.
For $K_0 < 2.4048$, the refractive index rises from its
initial value of 1, showing how the speed of light in the lattice
drops. At the zero of the Bessel function, the refractive 
index diverges, corresponding to the system becoming infinitely
optically dense. For larger values of $K_0$ the refractive index
becomes {\em negative}, indicating that in this regime (shaded grey)
negative refraction \cite{veselago} occurs.

\begin{figure*}
\begin{center}
\includegraphics[width=0.23\textwidth,angle=-90,clip=true]{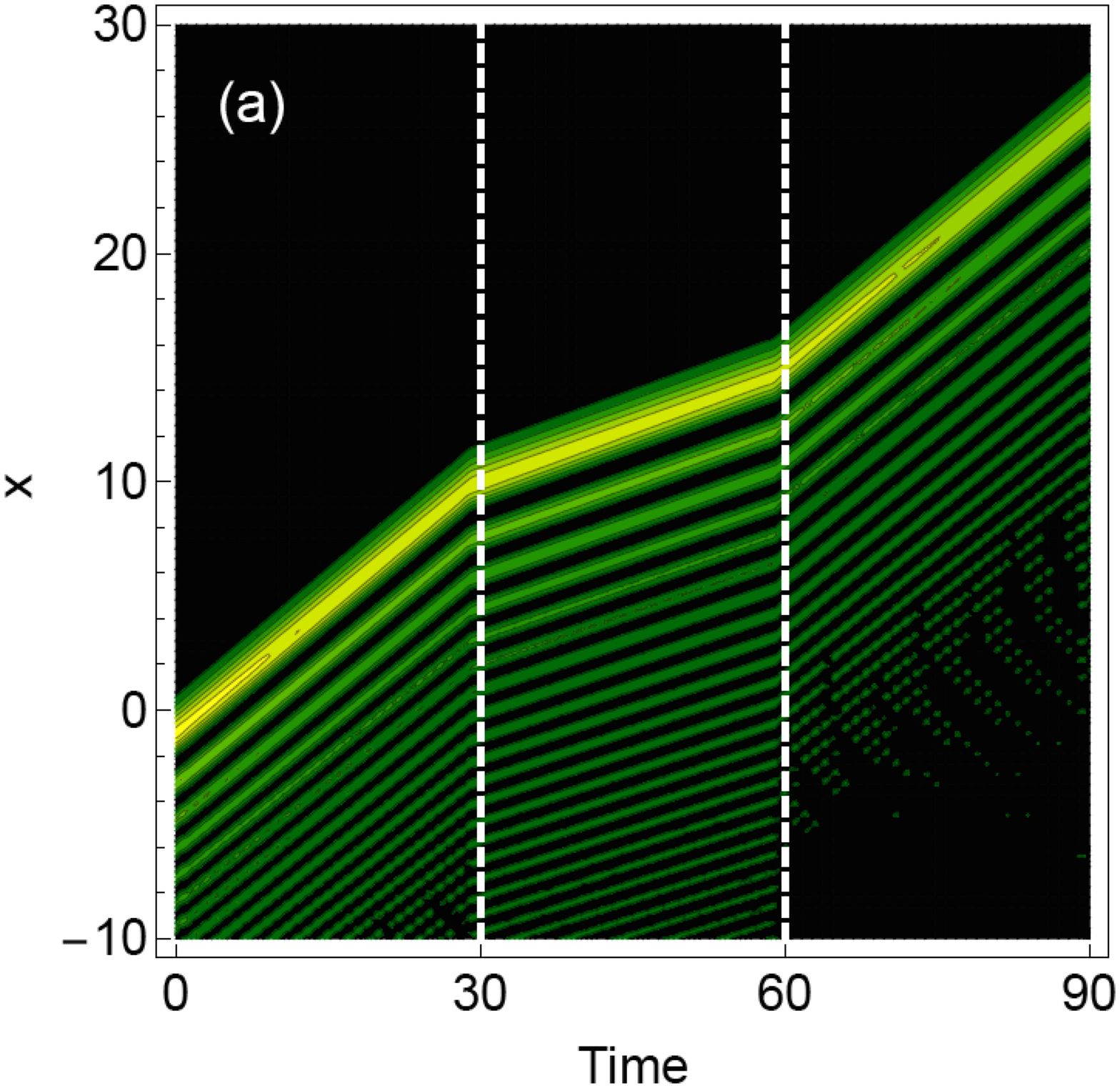}
\includegraphics[width=0.23\textwidth,angle=-90,clip=true]{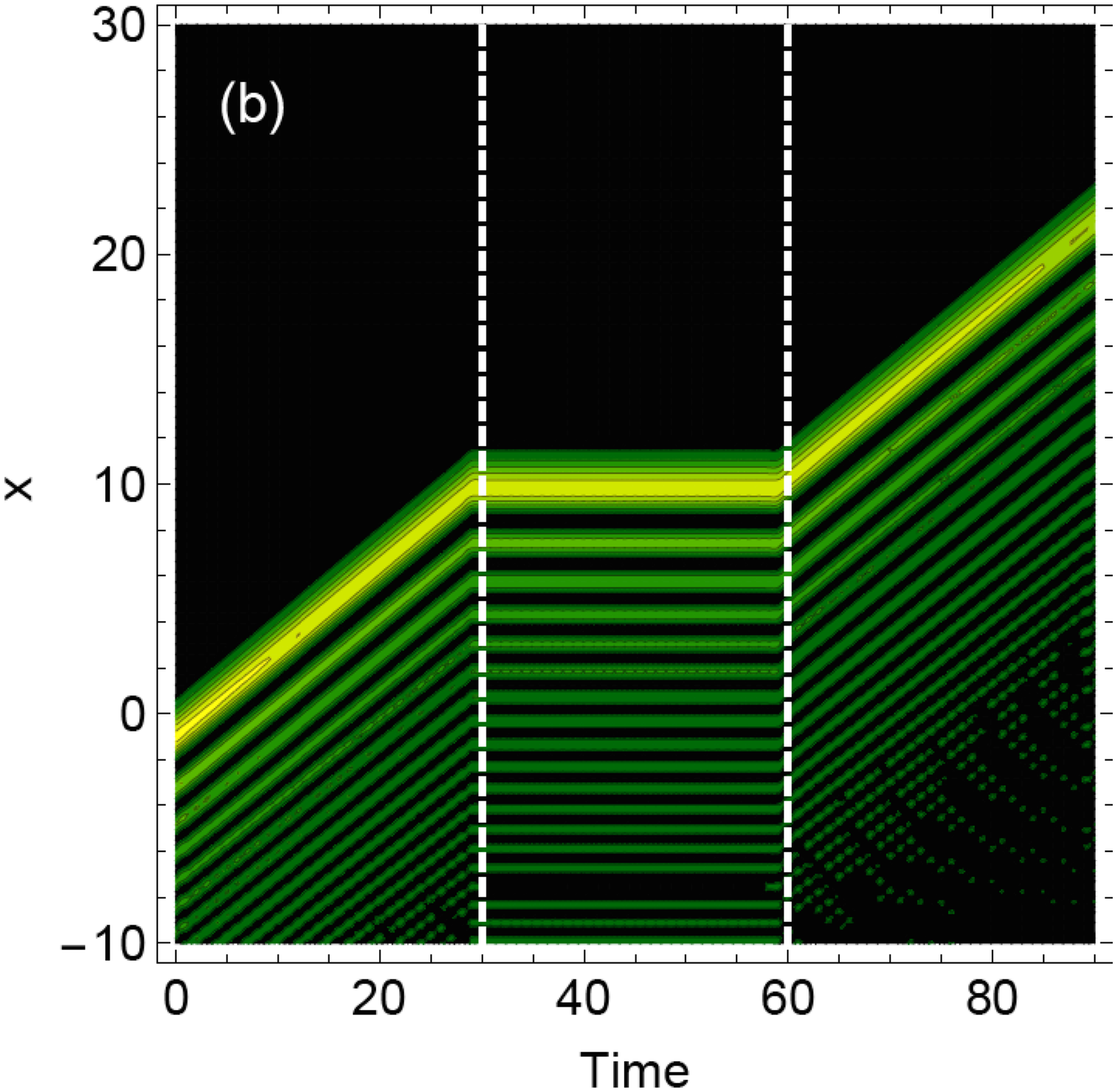}
\includegraphics[width=0.23\textwidth,angle=-90,clip=true]{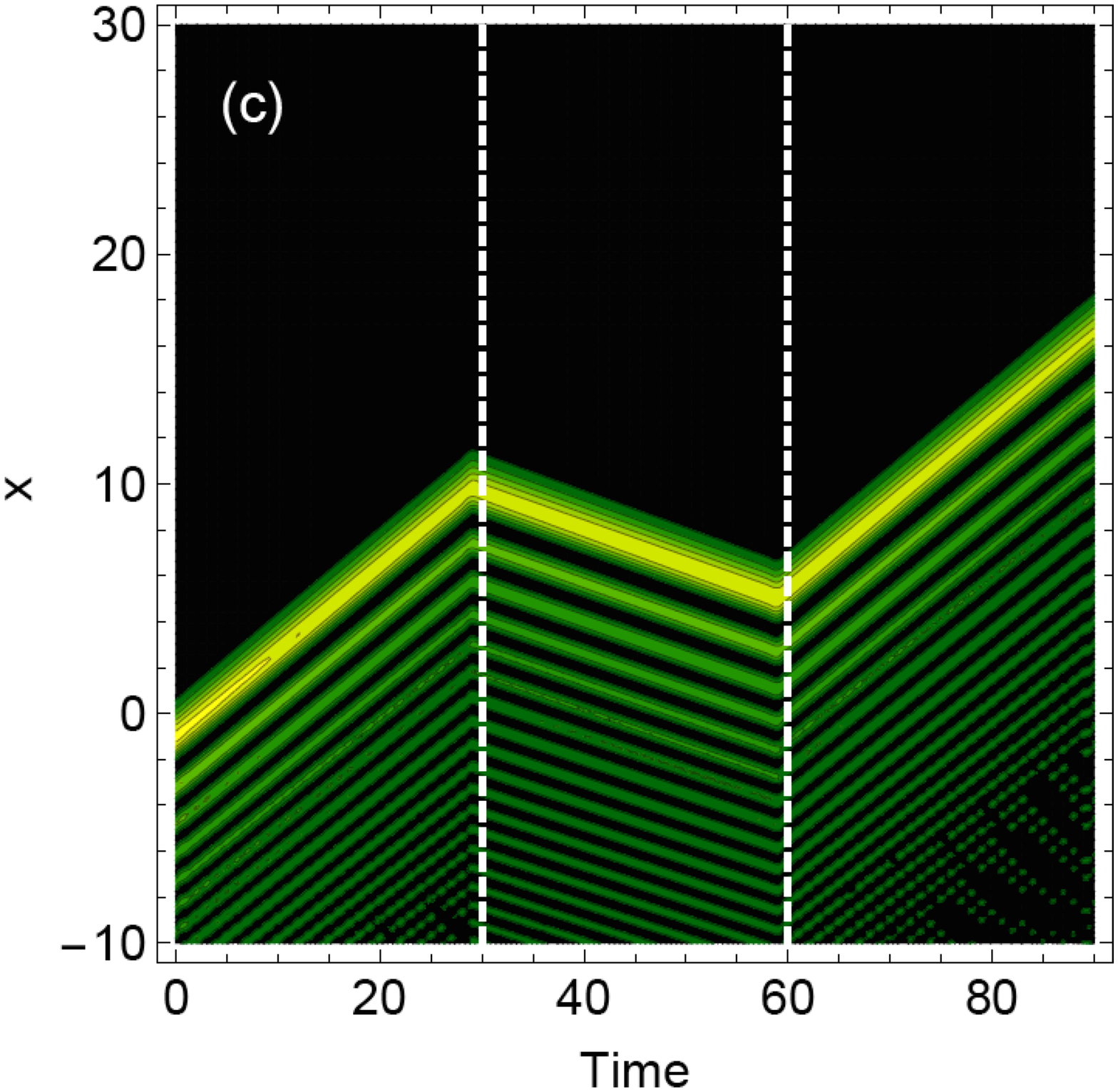}
\end{center}
\caption{Trajectories of an Airy wavepacket in a driven lattice,
simulation parameters: $\omega = 2 \pi$, $\Delta x = 0.2$.
In all cases $K_0$ is initially set to a value of
$K_0 = 0.5$, for which $\Jeff$ is slightly
reduced from its undriven value. At $t= 30$, $K_0$ is abruptly
changed to another value, while at $t = 60$ it reverts
to is original value; these boundaries are
marked by the vertical dashed lines.
(a) Between $t=30$ and $t=60$, $K_0$ is set to a value
of $K_0 = 1.691$. In this time interval the velocity
of the wavepacket is reduced by more than half, and the peaks 
appear to undergo refraction at the boundaries described by Snell's law.
(b) By tuning $K_0$ to a zero of the Bessel function ($K_0 = 2.4048$),
the wavepacket's motion is frozen. This corresponds to the medium's
refractive index diverging, producing an analogous effect to 
``stopped light''.
(c) Setting $K_0 = 3.80$ causes the wavepacket to reverse
its motion since the effective tunneling becomes
negative. This behaviour mimics the
phenomenon of negative refraction.}
\label{driven}
\end{figure*}

In Fig. \ref{driven} we show the probability densities for
Airy wavepackets under various driving conditions,
obtained by the numerical simulation of the full
time-dependent Hamiltonian (\ref{driving}). In all cases
the initial amplitude of the driving was set to $K_0 = 0.5$,
giving a refractive density of $1.07$, meaning that the
Airy wavepacket moved at a slightly lower velocity than in
the absence of driving. At $t=30$ the amplitude of the driving
is abruptly changed to a different value, and then restored to $K_0 = 0.5$
at $t=60$.

Fig. \ref{driven}a shows the result of reducing $\Jeff$
to a smaller, but positive value. It can be seen
that the trajectories of the peaks change their angle of
propagation with respect to the $t$-axis, and the peaks continue
moving along relativistic linear paths. The adjustment of
their velocity to the lower value of $\vmax$ appears to
occur essentially instantaneously, and does not cause any
appreciable deformation of the wavepacket's profile. When
$K_0$ is restored to its previous value, the original form of
propagation of the wavepacket resumes, with the trajectories
moving parallel to their original course. This behaviour
strongly resembles the standard refraction of light by a
slab of material with a positive refractive index.

In Fig. \ref{driven}b we show the effect of tuning $K_0$ to
a value of 2.404, close to the zero of the Bessel function. In this
case the refractive index diverges, and when the lattice is
driven at this value of $K_0$ the motion of the wavepacket
is completely frozen. This is analogous to the phenomenon of
slow or ``stopped'' light \cite{slow_light} previously
seen in ultracold gases by using electromagnetically induced
transparency to manipulate the refractive index. 
Finally, Fig. \ref{driven}c shows the 
behaviour of the wavepacket when the effective tunneling is
renormalized to a negative value. The motion of the peaks
now reverses while this condition is fulfilled, mimicking the
effect of a light ray traversing a region with negative
refractive index.

\section{Conclusions}
\begin{figure}[t]
\begin{center}
\includegraphics[width=0.5\textwidth,clip=true]{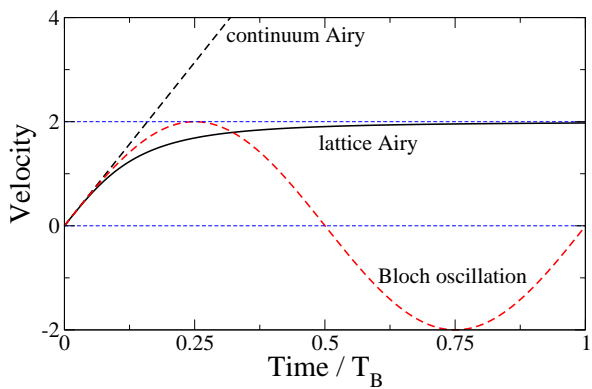}
\end{center}
\caption{Three different forms of behaviour for
accelerated quantum systems. The velocity of
a continuum Airy wavepacket rises linearly with time, and
increases without limit. In contrast, in a discrete lattice system,
the velocity initially rises linearly, but asymptotically approaches
the value of $\vmax$, which acts as the speed of light in
this system. It should be noted that in this case, the
proper acceleration of the wavepacket indeed stays constant.
Finally, tilting the lattice potential subjects a wavepacket
to a constant uniform force. The presence of the lattice, however,
means that the wavepacket does not accelerate uniformly,
but instead undergoes Bloch oscillation.}
\label{summary}
\end{figure}

We have studied the dynamics of an Airy wavepacket moving
in a lattice potential. Like its continuum counterpart \cite{berry}, the
lattice Airy wavepacket undergoes self-acceleration, but due
to the limitation on the maximum speed of propagation arising
from the lattice structure, this acceleration reduces in time
in accordance with relativistic kinematics.
This contrasts with the case of a particle on a lattice
subjected to a constant force. While the limited range
of velocity also plays a role in this case, the particle instead follows
an oscillatory motion -- Bloch oscillation -- in which its velocity
periodically cycles between $\pm \vmax$. We summarize
these three different forms of motion in Fig. \ref{summary}.

Lattice Airy wavepackets thus provide a means to observe
relativistic effects by lowering the effective speed of light
to a level which is rather convenient for experiment.
This opens avenues to simulate systems which are otherwise
rather resistant to physical realization, such as
the relativistic harmonic oscillator \cite{relativistic_ho}, as
well as employing relativistic effects to produce
more exotic effects such as enhancing the lifetimes of
unstable particles via time dilation \cite{dirac_airy}.
We have also shown how the trajectory of the lattice Airy wavepacket
can be manipulated by driving the lattice. This high level of
controllability, as opposed to the single
ballistic path \cite{ballistic} of the
continuum case, makes these wavepackets ideal candidates
to convey matter coherently from one point in a lattice to another,
with many possible applications to quantum information transfer \cite{bus}.
In the deep relativistic limit, in which the wavepacket's motion
is photon-like, this control over the trajectory can be
used to mimic a material with negative refractive index, and
could in the future be used to study perfect lensing \cite{pendry}
of matter waves.

Finally, we turn to possible experimental implementations of this system.
The driven lattice experiments of Ref.\cite{lignier} for example,
used a gas of approximately $10^5$ ultracold ${ }^{87}\mathrm{Rb}$ atoms,
held in an optical lattice with a well-spacing of 426 nm and
a tunneling frequency of $J \sim 100$Hz. 
This corresponds to an effective value of
the speed of light of $\vmax = 85 \ \mu\mathrm{m/s}$, 12 orders
of magnitude smaller than $c$ in free space.
If we take a lattice spacing
of $\Delta x = 0.2$, so that the first and second peaks of the
Airy wavepacket are separated by 12 lattice spacings, then
from Table \ref{values} we can see that $\alpha = 0.015$ in
lattice units. This translates to $64 \ \mu\mathrm{m} / \mathrm{s}^2$
in physical units, using these values for $d$ and $J$. Thus over
a time evolution of one second the wavepacket would move $\sim 75$ lattice
spacings, which should be easily resolvable using
quantum gas microscopy \cite{bloch_oscs,greiner,microscope}.
In Ref.\cite{wolf} it was noted that atomic interactions
did not effect the motion of the continuum wavepacket much,
except in the limit of very strong interactions for which it would
decay by shedding solitons. Accordingly we believe that for weak
to moderate interactions, the lattice Airy wavepacket should be
realizable in state-of-the-art experimental setups.
Including the effects of interactions and temperature,
and generalizing these results to higher dimensions, remain 
fascinating subjects for future research.

\acknowledgments
This work has been supported by Spain's MINECO through
Grant Nos. FIS2013-41716-P and FIS2017-84368-P.

\bibliographystyle{aipnum4-1}
\bibliography{airy_bib}

\end{document}